\documentclass[english]{article}
\usepackage{lmodern}

\usepackage[T1]{fontenc}
\usepackage[latin9]{inputenc}
\usepackage{color}
\usepackage{babel}
\usepackage{float}
\usepackage{fancybox}
\usepackage{calc}
\usepackage{amssymb}
\usepackage[unicode=true,
 bookmarks=true,bookmarksnumbered=false,bookmarksopen=false,
 breaklinks=false,pdfborder={0 0 1},backref=false,colorlinks=true]
 {hyperref}
\hypersetup{pdftitle={An Overview of Nominal-Typing versus Structural-Typing in Object-Oriented Programming},
 pdfauthor={Moez A. AbdelGawad},
 pdfsubject={Nominally-Typed OOP}}

\makeatletter

\newcommand{\noun}[1]{\textsc{#1}}

\newenvironment{lyxcode}
{\par\begin{list}{}{
\setlength{\rightmargin}{\leftmargin}
\setlength{\listparindent}{0pt}
\raggedright
\setlength{\itemsep}{0pt}
\setlength{\parsep}{0pt}
\normalfont\ttfamily}%
 \item[]}
{\end{list}}
\newcommand{\code}[1]{\texttt{#1}}

\makeatother

\begin{document}

\title{An Overview of Nominal-Typing versus Structural-Typing in Object-Oriented
Programming\\
(\emph{with code examples})}

\author{Moez A. AbdelGawad\\
\emph{moez@cs.rice.edu}}

\maketitle

\begin{abstract}
{\large{}}\global\long\def\NOOP{\mathbf{NOOP}}
$\NOOP$~\cite{NOOP,NOOPbook,NOOPsumm,AbdelGawad14} is a mathematical
model of nominally-typed OOP that proves the identification of inheritance
and subtyping in mainstream nominally-typed OO programming languages
and the validity of this identification~\cite{InhSubtyNWPT13}. This
report gives an overview of the main notions in OOP relevant to constructing
a mathematical model of OOP such as $\NOOP$. The emphasis in this
report is on defining nominality, nominal typing and nominal subtyping
of mainstream nominally-typed OO languages, and on contrasting the
three notions with their counterparts in structurally-typed OO languages,
\emph{i.e.}, with structurality, structural typing and structural
subtyping, respectively. An additional appendix demonstrates these
notions and other related notions, and the differences between them,
using some simple code examples. A detailed, more technical comparison
between nominal typing and structural typing in OOP is presented in
other publications (\emph{e.g.,}~\cite{AbdelGawad13}).
\end{abstract}

\section{Main Notions in OOP}

\subsection{Objects, Fields and Methods}

An object in object-oriented programming can be viewed as a ``service
provider'', \emph{i.e.}, as `an entity that provides a service'.
An object provides its service by providing a set of object fields,
as the inactive component of the object, and a set of object methods,
as its active component.

A \emph{field} of an object is a binding of a name to some other object.
The field name is used to access and interact with the object bound
to it. A \emph{method} of an object is a binding of a name to a function
that performs some computation when the method is invoked (on zero
or more objects, as \emph{method arguments}) and returns an object
as a result of the computation. The computation done by a method involves
accessing fields of some objects and/or calling their methods. The
object containing a method is always available as an implicit argument
to that method (\emph{e.g.}, under the special name \code{this} or
\code{self}), thereby providing access to the fields and methods
of the object inside the code for the method.

Collectively, the fields and methods of an object are called \emph{members}
of the object. The set of members of an object is finite. In statically-typed
OO languages this set is also fixed, and thus cannot change at runtime.
An object responds to a method call by performing the computation
associated with the method name, as implemented by the code of the
method, and it returns the result object. Collectively, the response
of an object to field accesses and method calls and the logical relation
between this response and the method arguments define the behavior
of the object. The behavior of an object defines the service the object
provides.

\subsection{Encapsulation}

Object-oriented programming is defined by two features: encapsulation
and inheritance. Because an object is viewed as a whole integrated
service provider, its members are not a collection of unrelated, independent
members. Members of an object usually mutually depend on each other.
Their interplay collectively provides the service of the object. Methods
of an object often call one another, where an object can recursively
invoke its own methods (via the special argument \code{self/this}).
Methods of an object access its fields to obtain information on the
``state'' of the object. The high dependency and coupling between
fields and methods of an object necessitates the two components (\emph{i.e.},
fields and methods) to be bound together (``embedded'') inside the
object. The embedding and binding together of the active and inactive
members of an object inside the object is expressed by stating that
objects \emph{encapsulate} their members. Further, encapsulation
in OOP not only refers to the bundling together of data and code that
accesses this data inside objects but also to allowing some implementation
details of an object (\emph{e.g.}, the code of its methods) and some
details of its representation and structure (\emph{i.e.}, some of
its fields and methods) to be hidden from other objects. This aspect
of OOP encapsulation is sometimes also called information hiding.
Information hiding helps maintain invariants of objects. It also allows
implementation and representation details of objects that are hidden
(\emph{i.e.}, not expressed in object interfaces, discussed below)
to change, without the changes impacting client code.

\subsection{Contracts, Classes, Class Names and Nominality}

Many objects in an OO program share similar behavior, have similar
properties, and provide the same service, only with some little variations\emph{
}(\emph{e.g.}, these objects may only differ in the exact values/objects
bound to their fields). A \emph{contract} is a set of formal or informal
statements that express common behavior and common properties of objects
shared by a set of similar objects. As an expression of the common
service provided by the set of similar objects, the statements of
a contract are assumed to hold true for all objects of the set. A
\emph{class} is a syntactic construct that OO languages offer for
the specification of the behavior of objects that provide more or
less the same service and that abide by and maintain the same contract.
A class is used as a template for the construction of these similar
objects. Objects produced using a certain class are called \emph{instances
}of that class.

A class has a name, called its \emph{class name}. Class names are
required to be unique in an OO program (using fully-qualified class
names, if necessary). A class name is always associated, even if informally,
with the common contract maintained by instances of the class. To
relate objects in class-based OOP to the semantic class contracts
the objects maintain, instances of a class have the name of the class
as well as class names of its superclasses as part of their identity.
That is, in class-based OOP \emph{class names inside objects are part
of what it means to be an object}. Class name information inside
an object is called its \emph{nominal information}. The association
of class names with contracts and the availability of nominal information
at runtime enables OO developers to design their software and control
its behavior based on the contracts of objects in their software%
\footnote{Using, for example, tests on class names, such as the \code{instanceof}
check in Java and the \code{isMemberOf} operation in Smalltalk. Nominally-typed
OO languages also do runtime checks akin to \code{instanceof} when
passing arguments to called methods and when returning objects returned
by the methods, based on method signatures.%
}.

To emphasize the fact that objects in class-based OOP have class names
as part of their meaning they are sometimes called \emph{nominal objects}.
A nominal object is always tied to the class (and superclasses) from
which it was produced, via the class name information (\emph{i.e.},
nominal information) embedded inside the object. Having class names
as part of the meaning of objects is called \emph{nominality}. An
OO language with nominal objects is a \emph{nominal} OO language.
Examples of nominal OO languages include Java~\cite{JLS05}, \noun{C\#~\cite{CSharp2007},
}Smalltalk~\cite{Smalltalk98}, \noun{C++}~\cite{CPP2011}, Scala~\cite{Odersky09},
and X10\noun{~}\cite{X1011}.

An OO language that does not embed class names in objects is
called a \emph{structural }OO language. The term `structural' comes
from the fact that an object in such a language is simply viewed
as a record containing fields and methods but with no class name information,
and thus no mention of the contracts maintained by the object. The
view of objects as records thus reflects only their structure. Examples
of structural OO languages include Strongtalk\noun{~}\cite{Bracha1993},
Moby\noun{~\cite{Fisher1999}}, PolyToil\noun{~}\cite{Bruce2003},
and OCaml~\cite{OCamlWebsite}.

\subsection{Shapes, Object Interfaces and Nominal Typing}

We call the set of names of members of an object the \emph{shape}
of the object. Object shapes are sets of labels, \emph{i.e.}, are
sets of names of members of the objects. Given that the shape of all
instances of a class in statically-typed mainstream OO languages is
fixed (\emph{i.e.}, is an invariant of instances of the class), we
also talk about shapes of classes.%
\footnote{We believe the shape of an object is a notion that is important and
intuitive enough to deserve a name of its own. If we need to emphasize
that fields and methods of objects have separate namespaces (\emph{e.g.},
as in Java), we then speak of the \emph{fields shape} and the \emph{methods
shape} of an object (or, of a class, a class signature, or a class
signature closure).%
}

The notion of encapsulation (as information hiding) in OOP motivates
the notion of object interfaces. An \emph{object interface} is an
informal notion that specifies, possibly incompletely, how an object
is viewed and should be interacted with by ``the outside world'',
\emph{i.e.}, by other objects of an OO program.%
\footnote{The interface of an object, thus, sort of tells the ``set of rules''
other objects have to follow to interact with the object. Despite
some similarity, the notion of an object interface should not be confused
with the more concrete notion of \code{interface}s that exists in
some OO languages.%
} It is hard to find a universally-accepted definition of the notion
of object interfaces. An object interface typically contains information
on the names of fields and methods of objects (their shapes), and
the interfaces of these members themselves. We adopt a nominal view
of object interfaces. In the nominal view of interfaces we adopt,
an object interface further includes the class name information of
the class from which the object is produced. Accordingly, the contracts
associated with names of classes are part of the public interface
of the object; these contracts are an essential part of how the object
should be viewed and interacted with by other objects. Class name
information inside an object interface can be used to derive other
information about an object, like its full class inheritance information.
This inheritance information is also part of how an object should
be viewed by other objects.

In class-based OOP, object interfaces are the basis for the formal
definition of class signatures and, in statically-typed OO languages,
are also the basis for the formal definition of \emph{class types}.
The nominality of object interfaces in mainstream OOP causes class
signatures, and class types, to be nominal notions as well. A statically-typed
nominal OO language where objects are further associated with class
types is thus called a \emph{nominally-typed} OO language.

Nominally-typed OO languages allow readily expressing circular (\emph{i.e.},
mutually-dependent) class definitions. The wide need for circular
class definitions and the ease by which recursive typing can be expressed
in nominally-typed OO languages is one of the main advantages of nominally-typed
OOP. According to Benjamin Pierce~\cite{TAPL}, ``The fact that
recursive types come essentially for free in nominal systems is a
decided benefit {[}of nominally-typed OO languages{]}''.

OO languages where types of objects are structural types (\emph{i.e.},
expressed as record type expressions that denote record types, with
no class name information) are called \emph{structurally-typed} OO
languages. See the appendix for examples of both nominal and structural
OO type expressions.

\subsection{Inheritance and Nominal Subtyping}

Inheritance is the second defining feature of OOP, where it is the
main mechanism for sharing and reuse. Inheritance is a syntactic relation
explicitly expressed and defined between classes of an OO program.
It allows the code of a class (including its field and method definitions)
to be defined in terms of the code of other classes. The defined
class is said to inherit from (or, extend) the other classes. Inheritance
is sometimes also called \emph{subclassing}, where the inheriting
class is called a \emph{subclass} while a class that is inherited
from is called a \emph{superclass}.

Two levels at which inheritance takes place need to be distinguished.
At the level of values (\emph{i.e.}, objects), inheritance involves
the sharing of code of object methods and fields (implementation inheritance).
At the type level, inheritance involves the sharing of interfaces
and contracts of objects. Given our goal of analyzing and improving
type systems of mainstream OO languages, our main focus is on type-level
inheritance. Unless otherwise noted, all references to inheritance
refer to type-level inheritance rather than object-level inheritance.

Inheritance in mainstream OOP is a nominal relation\emph{.} Inheritance
is specified between classes using class names in mainstream OO languages,
asserting that a subclass \emph{also inherits the class contracts}
associated with the names of its superclasses.%
\footnote{\textcolor{cyan}{}Similar to OO languages such as Java and C\#, we
allow multiple inheritance of \code{class}es/\code{interface}s.
The names of \code{class}es and \code{interface}s, alike, are always
associated with class contracts in mainstream OO languages. For example,
as part of its ``class'' contract, \code{interface} \code{Comparable}
in Java requires a total ordering on instances of any inheriting \code{class}.%
} Because of being explicitly specified, inheritance in class-based
OOP is always an intended relation between classes and is never an
implicit, accidental or ``spurious'' relation. Most importantly,
in nominally-typed mainstream OOP, a subclass that explicitly inherits
from some superclass is explicitly declaring that \emph{its instances
maintain the same contract} \emph{associated with the superclass}.

\section{\label{sub:Types-and-Type-Expressions}Types and Typing in OOP}

\textcolor{cyan}{}In computer programming, object-oriented or otherwise,
typing (\emph{i.e.}, the use of types and type systems) is mainly
a means to \emph{disallow improper use} of data values (``well-typed
programs can't go wrong''~\cite{MilnerPolymorphism78}). The \emph{type
soundness }(also called \emph{type safety}) of a programming language
is a statement about the language stating that ``well-typed'' programs
written using this language do not exhibit certain language-specific
undesirable program behaviors (\emph{i.e.}, programs that have no
\emph{type errors} do not ``go wrong''). Type-soundness is a desirable
property of programming languages. Much of PL research has been focused
on the study of type systems of programming languages to prove their
type safety.

The most common interpretation of types in programming languages is
that a type is \emph{a set of similar} \emph{values}. Under this set-theoretic
interpretation, types can be large sets of values that satisfy weak
similarity constraints, or be small sets of values that satisfy strong
similarity constraints, with other types filling the spectrum in-between
these two extremes.

The most desirable criteria for judging the similarity of software
values and data is to judge the similarity of their behavior and
of how these values should be used. Defining types according to the
similarity of behavior and usage of data values is called \emph{semantic
typing}.

Judging the similarity of behavior and usage of data values is known
to be computationally undecidable, generally-speaking. The intractability
of semantic typing motivates using \emph{syntactic typing}. Syntactic
typing dictates that program values are assigned syntactic expressions,
called \emph{type expressions}, that denote the (syntactic) type of
these values and phrases. Type expressions usually express properties
of values that can be easily checked, thereby making syntactic typing
tractable. The type corresponding to (or, denoted by) a given type
expression is the set of values that can be assigned the given type
expression as their (programmatic) type. In syntactic typing, the
similarity of the values of a type comes from the fact that all the
values have the same property expressed by the type expression denoting
the type.

\subsection{Class Types and Nominal Typing}

In the context of OOP, typing is a means to disallow improper use
of objects. Examples of improper use of objects include: (1) attempting
to access a non-existing field or to call a non-existing method of
an object, (2) allowing a field access or a method call for a field
or method not expressed in an object interface (\emph{i.e.}, breaching
the abstraction/information hiding offered by object interfaces),
or (3) not maintaining the contract associated with the class of an
object or the contracts associated with its superclasses.

Nominally-typed OOP languages use syntactic object features such
as inheritance relations and object interfaces to decide the similarity
of objects and to characterize their types. As mentioned earlier,
given that class names are part of object interfaces (\emph{i.e.},
are part of the public view of objects) and inheritance relations
are explicitly specified between class names, type expressions for
objects in nominally-typed OOP are nominal type expressions. The sets
of objects these type expressions denote are called \emph{class types}.
Class types are nominal notions, since class types with different
class names are different class types that denote different sets of
objects. In structurally-typed OOP, on the other hand, object type
expressions\emph{ }express a structural view of object interfaces
that does \emph{not} include class names nor inheritance relation
information. As such, type expressions of objects in structurally-typed
OOP are the same as record type expressions. The sets of objects these
type expressions denote are the same as \emph{record types} well-known
in the world of functional programming and among PL researchers.

In many mainstream OO languages such as Java and C\#, types of objects
are class types not record types, \emph{i.e.}, they are nominally-typed
not structurally-typed. Reflecting the nominality of class types in
nominally-typed OO languages, class names are used also as names for
class types in nominally-typed OOP. This fact demonstrates the central
role played by class names in the type systems of nominally-typed
mainstream OO languages. A crucial advantage of the nominality of
class types is that objects of a class type are not only similar structurally
but the objects of the class type (given the association of class
names with contracts) all \emph{maintain the same contract} associated
with the name of the class type. This property, which gets typing
in nominally-typed mainstream OOP closer to semantic typing, is a
main drive behind the adoption of nominal typing among OO developers
and language designers.

\subsection{Subtyping and Substitutability}

Given that objects in OOP can have different criteria for judging
similarity, and varying degrees of similarity can be used, the same
object can belong to multiple ``sets of similar objects'',\emph{
}according to the degree of similarity required of elements of the
sets and the criteria used to judge their similarity. If these sets
of similar objects are denotable by type expressions of the language,
the sets are types of the language (programmatic types). If values
of a programming language can be assigned multiple types the language
is said to support \emph{type polymorphism}. When some type is a subset
of (\emph{i.e.}, its elements are included in the elements of) another
type, the first type is said to be a \emph{subtype} of the second
type, which in turn is called a \emph{supertype} of the first type.
Because of subsumption, a value assigned a subtype automatically polymorphically
is assigned any supertype of the subtype. This form of type polymorphism
is called \emph{subtyping polymorphism}, or just \emph{subtyping}.
Subtyping has an intuitive counterpart in everyday lives, with plenty
of everyday examples---sometimes under the name of `inclusion' or
`containment'---that get paralleled in OO software.

All statically-typed OO languages support subtyping polymorphism.
In nominally-typed OOP, inheritance relations and object interfaces
(upon which the definition of type expressions of objects is based)
allow varying degrees of similarity between objects to be defined
in a natural way. When concrete expressions of object interfaces (\emph{e.g.},
class signatures) are interpreted as type expressions of objects,
object interfaces of classes that have more superclasses and more
object members are interpreted as requiring \emph{more} object-similarity
constraints. These richer object interfaces thus denote smaller\emph{
}sets of objects (\emph{i.e., }smaller class types).

\subsection[LSP and Nominal Subtyping]{The Liskov Substitution Principle and Nominal Subtyping}

Just as for typing, semantic subtyping is more desirable but less
tractable than syntactic subtyping. Semantic subtyping is commonly
expressed as the Liskov Substitution Principle (LSP), which is familiar
to many OO developers. The LSP states that in a computer program,
a type \code{S} is a subtype of a type \code{T} if and only if objects
of type \code{T} may be replaced with (\emph{i.e.}, substituted by)
objects of type \code{S}, without altering any of the main behavioral
properties of that program.

To achieve more precise subtyping, the inheritance relation between
classes and the subtyping relation between class types are completely
identified (are in one-to-one correspondence) in nominally-typed OOP~\cite{InhSubtyNWPT13}.
Accordingly, an OO language that puts nominal information, and thus
class contracts, in consideration while deciding the subtyping relation
is a \emph{nominally-subtyped} OO language (all known nominally-typed
OO languages are also nominally-subtyped). An OO language that does
not use nominal information while deciding the subtyping relation
is a \emph{structurally-subtyped} OO language. In nominally-typed
mainstream OO languages, given the association of class names with
class contracts, the inclusion of class names in deciding the subtyping
relation makes the subtyping relation semantically more precise (since
it incorporates more behavioral properties) than structural subtyping.

\section*{Acknowledgment}

The author would like to thank Kariem ElKoush for his comments and
feedback on earlier drafts of this report.

\bibliographystyle{plain}
\bibliography{oop-techoverview2}

\appendix
\global\long\def\ext{\blacktriangleleft}
\global\long\def\subsign{\trianglelefteq}

\part*{\label{chap:Code-Examples}Code Examples}

In this appendix we present code examples that concretely demonstrate
the concepts and notions discussed in this report. Unless otherwise
noted, code examples in this appendix use the syntax of \noun{Java}-like
OO languages. To simplify the inheritance and subtyping relations,
in the examples we do \emph{not }assume that all classes have to inherit
from a single superclass (\emph{e.g.}, class \code{Object}).

\section{\label{sec:Classes}Classes}

As a first example, we assume a declaration of class \code{Object}
as in Figure~\ref{fig:class-Object}.

\noindent 
\begin{figure}[H]
\noindent %
\shadowbox{\begin{minipage}[t]{1\linewidth}%
\begin{lyxcode}
\textbf{class}~Object~\{

~~//~Classes~with~no~explicit~constructors~are~always

~~//~assumed~to~have~a~default~constructor~that~simply

~~//~initializes~the~fields~of~the~constructed~object.

~

~~Boolean~equals(Object~o)\{

~~~~\textbf{return}~(o~\textbf{is}~Object);

~~~~//~Equivalent~to:~`o.getClass()~==~Object.class'

~~\}

\}\end{lyxcode}
\end{minipage}}

\caption{\label{fig:class-Object}Class \protect\code{Object}}
\end{figure}

In class \code{Object}, and in other classes declared below, we make
use of the standard class \code{Boolean}. We assume class \code{Boolean}
has values \code{true} and \code{false} (or equivalents) as its
instances, and that it supports (via its methods) standard boolean
operations on boolean values.

Next, we will make use of classes \code{A}, \code{B}, \code{C},
\code{D} and \code{E}, whose declarations are as presented in Figure~\ref{fig:classes-ABCDE}.
These five simple classes are not quite realistic and they serve no
purpose except demonstrating the concepts and notions we discuss.

\noindent 
\begin{figure}[H]
\noindent %
\shadowbox{\begin{minipage}[t]{1\linewidth}%
\begin{lyxcode}
\textbf{class}~A~\{~//~no~superclasses

\}

~

\textbf{class}~B~\textbf{extends}~A~\{

~~//~add~no~members

\}

~

\textbf{class}~C~\textbf{extends}~B~\{

~~D~foo(D~d)~\{~\textbf{return}~d;~\}

\}

~

\textbf{class}~D~\{~//~no~superclasses

~~A~bar()~\{~\textbf{return}~\textbf{new}~A();~\}

\}

~

\textbf{class}~E~\textbf{extends}~D~\{

~~A~meth()~\{~\textbf{return}~\textbf{new}~A();~\}

\}\end{lyxcode}
\end{minipage}}

\caption{\label{fig:classes-ABCDE}Classes \protect\code{A}, \protect\code{B},
\protect\code{C}, \protect\code{D} and \protect\code{E}}
\end{figure}

As more complex example, we also assume the declaration of class \code{Pair}
as presented in Figure~\ref{fig:class-Pair}.

\noindent 
\begin{figure}
\noindent %
\shadowbox{\begin{minipage}[t]{1\columnwidth}%
\begin{lyxcode}
\textbf{class}~Pair~\textbf{extends}~Object~\{

~~Object~first;

~~Object~second;

~

~~Boolean~fstEqSnd()\{

~~~~\textbf{return}~first.equals(second);

~~\}

~~Boolean~equalTo(Pair~p)\{

~~~~\textbf{return}~first.equals(p.first)~\&\&

~~~~~~~~~~~second.equals(p.second);

~~\}

~~Boolean~equals(Object~p)\{

~~~~if(p~\textbf{instanceof}~Pair)

~~~~~~~\textbf{return}~equalTo((Pair)p);

~~~~\textbf{return}~false;

~~\}

~~Pair~setFirst(Object~fst)\{

~~~~\textbf{return}~\textbf{new}~Pair(fst,~second);

~~\}

~~Pair~setSecond(Object~snd)\{

~~~~\textbf{return}~\textbf{new}~Pair(first,~snd);

~~\}

~~Pair~swap()\{

~~~~\textbf{return~new}~Pair(second,~first);

~~\}

\}\end{lyxcode}
\end{minipage}}

\caption{\label{fig:class-Pair}Class \protect\code{Pair}}
\end{figure}

\section{Shapes}

Assuming a single namespace for fields and methods, the shape of (instances
of) class \code{Object}, as declared in Figure~\ref{fig:class-Object},
is the set
\begin{lyxcode}
\{equals\}.
\end{lyxcode}
The shapes of classes \code{A}, \code{B}, \code{C}, \code{D}
and \code{E}, as declared in Figure~\ref{fig:classes-ABCDE}, are,
respectively, the sets
\begin{lyxcode}
\{\},~\{\},~\{foo\},~\{bar\},~\{bar,~meth\}
\end{lyxcode}
(Note that the shapes of instances of \code{A} and \code{B} are
the \emph{same} set, namely the empty set $\phi$=\code{\{\}}). The
shape of class \code{E} is a supershape of (\emph{i.e.}, a superset
of) the shape of class \code{D}, which in turn, is a supershape
of the shape of classes \code{A} and \code{B}.

The shape of class \code{Pair}, as declared in Figure~\ref{fig:class-Pair},
is the set
\begin{lyxcode}
\noindent \{equals,~first,~second,~fstEqSnd,~equalTo,~setFirst,~\\
~setSecond,~swap\}.
\end{lyxcode}
The shape of class \code{Pair} is a supershape of the shape of class
\code{Object}%
\footnote{Note that because class \code{Object} has no fields, all its instances
are mathematically-equivalent. Mathematically-speaking, thus, class
\code{Object} has only one instance.%
}.

\noindent

\section{\label{sec:Record-Type-Expressions}Record Type Expressions}

A record type expression of class \code{Object}, expressing a structural
view of the object interface for instances of \code{Object}, is
\begin{lyxcode}
OSOI~$\triangleq$\textbf{~record\_type~}$\mu$O.~\{

~~B~equals(O)

\}~\textbf{and}~$\mu$B.~\{...~member~interfaces~of~class~Boolean...\}
\end{lyxcode}
where $\mu$ is the recursive types operator, and \texttt{\textbf{and
}}defines mutually-recursive types. (See the discussions in~\cite{TAPL,NOOPbook,AbdelGawad13}
on circularity in OOP.)

Record type expressions of classes \code{A}, \code{B}, \code{C},
\code{D} and \code{E}, respectively, are
\begin{lyxcode}
ASOI~$\triangleq$~\textbf{record\_type}~\{\}

BSOI~$\triangleq$~\textbf{record\_type}~\{\}

//~note~that~ASOI~and~BSOI~are~the~same.

DSOI~$\triangleq$~\textbf{record\_type}~\{

~~BSOI~bar()

~~//~Note~the~need~to~include~the~full~record~type~expression.

~~//~BSOI~is~a~``macro''.~~Even~though~class~A~is~used~to~define

~~//~class~D,~BSOI~is~used,~rather~than~ASOI,~to~make~a~point:

~~//~Names~of~record~type~expressions~are~just~``macro~names''.

~~//~The~names~can~be~changed~without~changing~the~meaning~of

~~//~the~defined~record~type~expressions.

\}

CSOI~$\triangleq$~\textbf{record\_type}~\{

~~DSOI~foo(DSOI)

\}

ESOI~$\triangleq$~\textbf{record\_type}~\{

~~ASOI~bar(),~

~~ASOI~meth()

~~//~Because~of~structurality,~BSOI~or~the~equivalent

~~//~unfolded~expression~`\textbf{record\_type}~\{\}'~could~be~used

~~//~in~place~of~ASOI~everywhere~ASOI~is~used.

\}
\end{lyxcode}

A record type expression expressing a structural view of the object
interface for instances of class \code{Pair} is
\begin{lyxcode}
PSOI~$\triangleq$\textbf{~record\_type~}$\mu$P.~\{

~~B~equals(O),~//~with~no~rebinding~from~O~to~P

~~O~first,~O~second,

~~B~fstEqSnd(),

~~B~equalTo(P),

~~P~setFirst(O),

~~P~setSecond(O),

~~P~swap()

\}~\textbf{and~}$\mu$O.\textbf{~}\{...member~interfaces~of~class~Object...\}

~~\textbf{and~}$\mu$B.~\{...member~interfaces~of~class~Boolean...\}
\end{lyxcode}

\section{\label{sec:Structural-Subtyping}Structural Subtyping}

The following record type expressions, from Section~\ref{sec:Record-Type-Expressions},
are in the structural subtyping relation, $<:$.
\begin{lyxcode}
BSOI~<:~ASOI~(and~spuriously~ASOI~<:~BSOI,~because~ASOI~=~BSOI)

CSOI~<:~BSOI~(a~genuine~``is-a'')

DSOI~<:~BSOI~(spurious~subtyping.~unwarranted~``is-a'')

ESOI~<:~DSOI~(a~genuine~``is-a'')

OSOI~<:~BSOI~(spurious~subtyping.~unwarranted~``is-a'')

PSOI~<:~OSOI~(a~genuine~``is-a'')
\end{lyxcode}

Note that a pair in the structural subtyping relation could express
a genuine ``is-a'' relation or an unwarranted, accidental relation
between instances of classes. For example,
\begin{lyxcode}
\textbf{record\_type}~\{\}~<:~\textbf{record\_type}~\{\}
\end{lyxcode}
which we expressed above, in disguise, as \code{BSOI~<:~ASOI} (and
\code{ASOI~<:~BSOI}), intuitively holds true when in reference to
objects of class \code{B} being also objects of class \code{A}.
This is something the developer (of class \code{B}) intended. It
is thus a genuine is-a relation. The same relation does not hold true,
however, when in it refers to objects of class \code{A} being objects
of class \code{B}. Viewing objects of \code{A} as objects of \code{B}
may not have been intended by the developer of class \code{A}. It
is an accidental (``spurious'') is-a relation. It is only a result
of the fact that \emph{structural subtyping does not capture the full
intention of class developers}.%
\footnote{Another example is the example of sets and multisets. Mathematically,
every set is a multiset, but a multiset that has repeated elements
is not a set. Classes modeling sets and multiset may have their structural
types (\emph{i.e.}, record type expressions) not reflecting the contracts
associated with the names `set' and `multiset' (of disallowing and
allowing repetitions, respectively). Structural OO implementations
of sets and multisets may thus allow instances of the two classes
modeling sets and multisets to be mixed, in particular allowing multisets
to be incorrectly viewed and interacted with as sets!%
}

The subtyping pairs \code{CSOI~<:~BSOI}, \code{ESOI~<:~DSOI} and
\code{PSOI~<:~OSOI} express genuine is-a relations when referencing
objects of classes \code{C} being \code{B}'s, \code{E}'s being
\code{D}'s, and \code{Pair}'s being \code{Object}'s, respectively.
The pairs \code{DSOI~<:~BSOI} and \code{OSOI~<:~BSOI} express an
unwarranted is-a relation when referencing objects of \code{D} being
\code{B}'s, and of \code{Object} being \code{B}'s.

\section{Signatures and Subsigning}

Based on declarations in Section~\ref{sec:Classes}, the signature
of class \code{Object}, expressing a nominal view of the interface
of \code{Object}, is
\begin{lyxcode}
\textbf{$Obj\triangleq$~sig~}Object~\{

~~equals:~Object$\rightarrow$Boolean

\}
\end{lyxcode}
and the signatures of classes \code{A}, \code{B}, \code{C}, \code{D}
and \code{E}, expressing nominal views of the interfaces of these
five classes, are
\begin{lyxcode}
\textbf{$A\triangleq$~sig}~A~\{\}

\textbf{$B\triangleq$~sig}~B~\textbf{ext}~A~\{\}

\textbf{$C\triangleq$~sig}~C~\textbf{ext}~B~\{

~~foo:~D$\rightarrow$D

\}

\textbf{$D\triangleq$~sig}~D~\{

~~bar:~()$\rightarrow$A

\}

\textbf{$E\triangleq$~sig}~E~\textbf{ext}~D~\{

~~bar:~()$\rightarrow$A,

~~meth:~()$\rightarrow$A

\}
\end{lyxcode}

\noindent Also, the signature of class \code{Pair} is
\begin{lyxcode}
\textbf{$Pair\triangleq$~sig~}Pair~\textbf{ext}~Object~\{

~~equals:~Object$\rightarrow$Boolean,

~~first:~Object,

~~second:~Object,

~~fstEqSnd:~()$\rightarrow$Boolean,

~~equalTo:~Pair$\rightarrow$Boolean,

~~setFirst:~Object$\rightarrow$Pair,

~~setSecond:~Object$\rightarrow$Pair,

~~swap:~()$\rightarrow$Pair

\}
\end{lyxcode}

(It should be noted that the syntax used to present examples of class
signatures above is different from the syntax generated by the more
mathematically-oriented abstract syntax rules for signatures that
are presented with $\NOOP$~\cite{NOOPsumm}. Even though equally
informative, the syntax of signatures we used here is closer to the
concrete syntax of classes that most mainstream OO developers are
familiar with.)

\noindent Next, we define the following signature environments and
signature closures
\begin{lyxcode}
\noindent $ObjectSE$~=~\{~$Obj$,~$Bool$~\}

~~~//~where~$Bool$~is~the~class~signature~of~class~Boolean

$Ase$~=~\{~$A$~\}

$Bse$~=~\{~$A$,~$B$~\}

$Cse$~=~\{~$A$,~$B$,~$C$~\}

$Dse$~=~\{~$A$,~$D$~\}

$Ese$~=~\{~$A$,~$D$,~$E$~\}

$PairSE$~=~\{~$Obj$,~$Bool$,~$Pair$~\}

~

$ObjSC$~=~(Object,~$ObjSE$)

$Asc$~=~(A,~$Ase$)$,$~$Bsc$~=~(B,~$Bse$)$,$~$Csc$~=~(C,~$Cse$)

$Dsc$~=~(D,~$Dse$)$,$~$Esc$~=~(E,~$Ese$)

$PairSC$~=~(Pair,~$PairSE$)
\end{lyxcode}

\noindent By the rules for signature environment extension we have
\begin{lyxcode}
$Bse$~$\ext$~$Ase$~(but~$Ase$~$\not\ext$~$Bse$)

$Cse$~$\ext$~$Bse$

$Ese$~$\ext$~$Dse$

$PairSE$~$\ext$~$ObjSE$
\end{lyxcode}
and by the rules of subsigning (See, for example, \cite[Sec. 4.5]{NOOPsumm})
we have
\begin{lyxcode}
$Bsc$~$\subsign$~$Asc$~(but~$Asc$~$\not\subsign$~$Bsc$)

$Csc$~$\subsign$~$Bsc$

$Esc$~$\subsign$~$Dsc$

$PairSC$~$\subsign$~$ObjSC$
\end{lyxcode}
Note that pairs in subsigning relation only express \emph{genuine}
``is-A'' relations. In particular, unlike we had for structural
subtyping (\emph{e.g.}, in Section~\ref{sec:Structural-Subtyping}),
for subsigning we do not have
\begin{lyxcode}
$Dsc$~$\subsign$~$Bsc$~(unwarranted~by~rules~of~subsigning,

~~~~~~~~~~~~since~$Dse\not\ext Bse$)

~

$Dsc$~$\subsign$~$Asc$~(unwarranted~by~rules~of~subsigning,~even~though

~~~~~~~~~~~~$Dse\ext Ase$,~since~A~$\notin super\_sigs(D)=\{\}$)

~

$ObjSC$~$\subsign$~$Bsc$~(unwarranted~by~rules~of~subsigning,

~~~~~~~~~~~~~~~since~$ObjSE\not\ext Bse$)
\end{lyxcode}

\noindent Using the class declarations presented in Section~\ref{sec:Classes},
the reader is invited to construct more examples of signature closures
in and outside the subsigning relation. Unlike structural subtyping
($<:$), the examples for subsigning demonstrate that \emph{subsigning,}
\emph{and thus nominal subtyping, fully captures the intention} \emph{of
class developers}.

\end{document}